\documentclass[11pt]{article}
\usepackage{amsmath,amssymb,amsfonts,amstext}
\usepackage{tabularx}
\usepackage{hyperref}
\usepackage{epsf}
\usepackage{graphicx}
\usepackage{amsfonts}
\usepackage{subfigure}
\usepackage[usenames]{color}
\definecolor{Blue}{rgb}{0,0.08,0.45}
\definecolor{Magenta}{cmyk}{0.1,0.8,0,0.1}
\definecolor{Orange}{rgb}{1,0.5,0}

\begin{document}

\title{\hfill\textbf{\small OUTP-09-05P}\\
Hybrid natural inflation from non Abelian discrete symmetry}
\author{ Graham G Ross \\
{\normalsize \textit{Rudolf Peierls Centre for Theoretical Physics,} }\\
{\normalsize \textit{University of Oxford, 1 Keble Road, Oxford, OX1 3NP, UK}%
}\\
\\
Gabriel Germ\'an\\
{\normalsize \textit{Instituto de Ciencias F\'isicas} }\\
{\normalsize \textit{Universidad Nacional Aut\'onoma de M\'exico,}}\\
{\normalsize \textit{Apdo. Postal 48-3, 62251 Cuernavaca Morelos, M\'exico}}}
\date{}
\maketitle

\begin{abstract}
A spontaneously broken global discrete symmetry may have pseudo Goldstone
modes associated with the spontaneous breaking of the approximate continuous
symmetry of the low dimension terms in the Lagrangian. These provide natural
candidates for an inflaton that can generate slow roll inflation. We show
that, in the case of a non Abelian discrete symmetry, the pseudo Goldstone
modes readily couple to further scalar fields in a manner that the end of
inflation is determined by these additional scalar fields, generating hybrid
inflation. We give a simple parameterisation of the inflationary potential
in this case, determine the inflationary parameters resulting, and show that
phenomenological successful inflation is possible while keeping the scale of
symmetry breaking sub-Plankian. Unlike natural inflation the inflation scale
can be very low. We construct two simple hybrid inflation models, one non
supersymmetric and one supersymmetric. In the latter case no parameters need
be chosen anomalously small.
\end{abstract}

\section{Introduction}

Field theory inflationary models typically suffer from the \textquotedblleft 
$\eta $\textquotedblright\ problem which arises from the need to keep the
inflaton very light, lighter than the mass scale set by the Hubble constant
during inflation. To solve this problem, including the effects of radiative
corrections, it is necessary to have an underlying symmetry which, when
exact, forbids the inflaton mass. The mass arises through the (small) terms
breaking the symmetry. Radiative corrections to the mass are then under
control because, due to the symmetry, they must be proportional to the
(small) terms breaking the symmetry. The only symmetries capable of
forbidding a scalar mass are a Goldstone symmetry, associated with the
spontaneous breaking of a continuous global symmetry, and a supersymmetry in
which the scalar mass is related to fermion mass and the latter can be
forbidden by a chiral symmetry.

A particularly attractive model based on a Goldstone symmetry is \textit{%
Natural Inflation} \cite{Freese:1990rb}. The Goldstone boson, $\phi ,$
arises due to the spontaneous breaking of a Global symmetry at a scale $f$.
However the symmetry is broken at the quantum level through an anomaly and
the would be Goldstone boson is a pseudo Goldstone boson acquiring a mass
through the anomaly. The resulting inflationary potential for the inflaton $%
\phi $ is given by 
\begin{equation}
V=V_{0}\left( 1+\cos \left( \frac{\phi }{f}\right) \right) .
\end{equation}%
The potential has only two parameters and describes the period at which the
density perturbations are produced until the end of inflation at which the
potential vanishes. The inflationary parameters resulting from this
potential are consistent with observation and the model makes definite
predictions for the tensor to scalar perturbations and the running of the
spectral index \cite{Freese:1991rb}.

Given the limited ways available to solve the $\eta$ problem it is of
interest to explore more general forms of the inflationary potential
resulting from a pseudo Goldstone inflaton\cite{Shaun:2008}. Here we
construct a class of models which give rise to hybrid natural inflation in
which a second field is responsible for ending inflation and setting the
cosmological constant to zero. The form of the inflationary potential when
the density perturbations are produced is given by

\begin{equation}
V=V_{0}\left( 1+a\cos \left( \frac{\phi }{f}\right) \right) ,  \label{hni}
\end{equation}%
where $a$ is a constant. As we shall discuss the appearance of the parameter 
$a,$ and the fact that the end of inflation is set by a second hybrid field,
can significantly change the expectations for the inflationary parameters.
Moreover it is possible to generate acceptable inflation for a lower
symmetry breaking scale than is the case for the original natural inflation.
This means one may have $f<M,$ where $M$ is the reduced Planck scale, $%
M=2.44\times 10^{18}$~GeV, so that higher dimension corrections to the
inflationary potential are under control. In addition the scale of inflation
is not determined and can be much lower than the upper bound set by the
non-observation of tensor fluctuations.

The models implementing hybrid natural inflation are based on non Abelian
discrete symmetries\footnote{%
Such symmetries often arise in string compactification \cite%
{Kobayashi:2006wq}.}. Such discrete symmetries can be equivalent to
continuous symmetries for the low dimension terms in the Lagrangian, the
approximate continuous symmetry being explicitly broken only by (small)
higher dimension terms. If the symmetry is spontaneously broken\ there will
be a light pseudo Goldstone boson that may be identified with the inflaton.
The non Abelian structure of the discrete symmetry is important in
constructing the hybrid nature \cite{Linde:1994} of the model because it\ allows for a
non-trivial coupling between the pseudo Goldstone inflaton field and the\
field responsible for ending inflation. We are aware of two earlier papers
which have explored this idea \cite{Stewart:2000pa},\cite{cohn:2000}
although they do not emphasize the general structure of Eq.(\ref{hni}). A
more recent paper \cite{Antusch:2008gw} explores the possibility that the
fields responsible for breaking a non Abelian discrete family symmetry
should be inflaton(s) but does not identify the pseudo Goldstone modes as
the candidate inflatons.

Here we present two simple models capable of giving the form of Eq.(\ref{hni}%
). The first is not supersymmetric and requires that the coefficient of the
leading term breaking the approximate continuous symmetry should be
anomalously small. It may be that such smallness follows if the non Abelian
discrete symmetry arises from the spontaneous breaking of an underlying
continuous symmetry, but we have not been able to demonstrate this yet. The
second example implements the non Abelian discrete symmetry in a
supersymmetric theory. The additional constraints of supersymmetry allows
for the construction of a viable model with no parameter chosen to be
anomalously small.

In Section 2 we determine the implications for the inflationary parameters
that result from the hybrid natural inflationary form of Eq.(\ref{hni}).
Sections 3 and 4 present the non supersymmetric and supersymmetric hybrid
natural inflation models respectively. Finally in Section 5 we present our
conclusions.

\section{Inflationary parameters from hybrid natural inflation}

It is straightforward to determine the constraints on the parameters of the
potential of Eq.$\left( \ref{hni}\right) $ following from the requirement of
acceptable inflation. This then leads to a determination of the detailed
form of the inflation\ observables produced by this potential. The
observables are given in terms of the usual slow-roll parameters \cite%
{Liddle:2000cg} 
\begin{eqnarray}
\epsilon &\equiv &\frac{M^{2}}{2}\left( \frac{V^{\prime }}{V}\right) ^{2}\simeq%
\frac{M^{2}}{2}\left( \frac{a}{f}\right) ^{2}\sin ^{2}\left( \frac{\phi }{f}%
\right),  \label{eps} \\
\eta &\equiv &M^{2}\frac{V^{\prime \prime }}{V}\simeq-M^{2}\left( \frac{a}{f^{2}}%
\right) \cos \left( \frac{\phi }{f}\right),  \label{eta} \\
\xi &\equiv &M^{4}\frac{V^{\prime }V^{\prime \prime \prime }}{V^{2}}\simeq-\left( 
\frac{M}{f}\right) ^{4}a^{2}\sin ^{2}\left( \frac{\phi }{f}\right)=-2\left(\frac{M}{f}\right)^2\epsilon. \label{garabato}
\end{eqnarray}%
In terms of these the tensor to scalar ratio is 
\begin{equation}
r=16\epsilon .
\label{tensor}
\end{equation}%
The spectral index is given by 
\begin{equation}
n_{\mathrm{s}}=1+2\eta -6\epsilon ,  \label{ns}
\end{equation}%
and the `running' of the spectral index characterised by $n_{\mathrm{r}}=%
\mathrm{d}n_{\mathrm{s}}/\mathrm{d}\ln k$ is 
\begin{equation}
n_{\mathrm{r}}=16\epsilon \eta -24\epsilon ^{2}-2\xi .  \label{run}
\end{equation}%
The density perturbation at wave number $k$ being 
\begin{equation}
\delta _{\mathrm{H}}^{2}(k)=\frac{1}{150\pi ^{2}}\frac{V_{\mathrm{H}}}{%
\epsilon _{\mathrm{H}}M^{4}}.  \label{perts}
\end{equation}
The subscript $\mathrm{H}$ above signals the scale at which observable 
inflation starts, some $\approx 60$ e-folds before the end of inflation when the
scalar density perturbation on the scale of our present Hubble radius~\footnote{ Numerically this is 
$H_0^{-1} \simeq 3000h^{-1}$~Mpc, where $h \equiv H_0/100$~km~s$^{-1}$~Mpc$^{-1} \sim 0.7$ 
is the Hubble parameter. The density perturbation can be measured down to $\sim 1$~Mpc, 
a spatial range corresponding to just 8 e-folds of inflation.} was generated.

What are the predictions for the inflationary parameters following from the
form of Eq.(\ref{hni})?

\subsection{Spectral index and tensor to scalar ratio.}

Let us start with the condition that the slow roll parameters are small.
Since the end of inflation is determined by the hybrid sector there is no
reason to expect $\cos \left( \frac{\phi }{f}\right) $ to be anomalously
small. In this case Eq.(\ref{eta}) requires that $a\left( \frac{M}{f}\right)
^{2}\ll 1$. If we further require $f<M$ so that the higher order
gravitational corrections to the potential are small we need $a\ll 1$ and
hence $\epsilon \ll \eta $ corresponding to a small tensor to scalar ratio.
The observed small value for $1-n_{\mathrm{s}}$ for $\cos \left( \frac{\phi }{f}\right)
\approx O(1), $ can be obtained simply by fixing $a$
\begin{equation}
a\approx  \frac{1}{2}(1-n_{\mathrm{s}})\left( \frac{f}{M} \right)^2.  \label{ans}
\end{equation}
From Eqs.(\ref{tensor}) and (\ref{eps}) follow an upper bound for the tensor
\begin{equation}
r<2 \left(1-n_{\mathrm{s}}\right)^2 \left( \frac{f}{M} \right)^2. \label{tensorineq}
\end{equation}
For example, with $f=0.1M$ and for $\cos \left( \frac{\phi }{f}\right)
\approx O(1), $ we have $a \approx 2\times 10^{-4}$
corresponding to $r<5.6\times 10^{-5}$, where we have used the value $n_{\mathrm{s}} > 0.947$ \cite {Komatsu:2008}. Smaller values of $f$ lead to much
smaller values of $a.$ The only way a significant value for $r$ can be
obtained is for $f>M$ corresponding to $a=O(1)$ as is the case for natural
inflation. However this is in the region where one may expect significant
higher order gravitational corrections to the potential. For $n_{\mathrm{s}} > 0.947$ (the lowest bound on $n_{\mathrm{s}}$ given in \cite {Komatsu:2008})  and $f$ taking its maximum value $f=M$ the absolute bound on $r$ is $5.6\times 10^{-3}$.

\subsection{Spectral index running.}

The running of the spectral index is dominated by the third term in Eq.(\ref{run}) and is given by 
\begin{equation}
n_{\mathrm{r}} \approx \frac{1}{4}\left( \frac{M}{f} \right)^2r, \label{runsol}
\end{equation}%
and from Eq.(\ref{tensorineq}) it follows that
\begin{equation}
n_{\mathrm{r}}<\frac{1}{2}\left(1-n_{\mathrm{s}}\right)^2. \label{runineq}
\end{equation}%
The inequality above $n_{\mathrm{r}}<1.4\times 10^{-3}$, is stronger than the current observational constraint $n_{\mathrm{r}}<0.019$ as given by \cite {Komatsu:2008}, \cite {Dunkley:2008}.

\subsection{Scale of inflation.}

The scale of inflation is set by the requirement that the density
perturbations have the observed magnitude. From Eqs.(\ref{perts}) and (\ref{eta}) we have 
\begin{eqnarray}
V_{\mathrm{H}}\approx V_{0} &\simeq &M^{4}A_{\mathrm{H}}^{2}\left( \frac{M}{f}\right)
^{2}a^{2}\sin ^{2}\left( \frac{\phi _{\mathrm{H}}}{f}\right) =\frac{1}{8}M^4A_{\mathrm{H}}^{2}r ,
\label{scale}
\end{eqnarray}%
where $A_{\mathrm{H}}\equiv\sqrt{75}\pi \delta _{\mathrm{H}}\approx \sqrt{75}\pi \left( 1.91\times 10^{-5}\right) .$ 
From Eq.(\ref{tensorineq}) we get a bound for the inflationary scale 
\begin{equation}
\Delta < \sqrt{\frac{ A_{\mathrm{H}}}{2}}\left( 1-n_{\mathrm{s}} \right)^{1/2}\left(\frac{f}{M}\right)^{1/2}M. \label{scalebound}
\end{equation}%
For $n_{\mathrm{s}} > 0.947$ we get $\Delta <  9 \times 10^{15}\left(\frac{f}{M}\right)^{1/2} GeV .$
This can nearly saturate the bound $V_{0}<\left( 10^{16}GeV\right) ^{4}$ \cite {Lyth:1996im} but typically is much less and can be very low. We shall illustrate this when we discuss two specific models of hybrid natural
inflation.

\subsection{Constraints from cosmological observations.}

Putting all this together, in Table \ref{table:1} we compare the predictions of \textit{Hybrid Natural Inflation} with those of \textit{Natural Inflation} and with recent observational bounds coming from the five years data of the WMAP team \cite{Komatsu:2008}, \cite{Dunkley:2008} and those from BAO \cite{Percival:2007} and SN surveys \cite{Riess:2004}, \cite{Astier:2006}, \cite{Woods-Vasey:2006}. 

\begin{table}[tbp] \small
\begin{center}
\begin{tabular}{|@{}c@{}|@{}c@{}|@{}c@{}|@{}c@{}|@{}c@{}|}\hline
 & WMAP5 & WMAP5+BAO+SN & Natural Inflation & Hybrid Natural Inflation  \\ \hline
$n_\mathrm{s}$ & $0.963^{+0.014}_{-0.015}$ & $0.960\pm 0.013$ & $\approx \lbrace^{1-  \left(\frac{M}{f}\right)^2,\quad f \leq 4M} _{1-\frac{2}{N_e},\quad f \geq 10M}$ & $\approx 1-2a\left(\frac{M}{f}\right)^2$ \\ \hline
$r$ &  $<0.43 (95\%CL)$ & $<0.22 (95\%CL)$ & $ < \frac{8}{N_e}< 0.2$ & $<2\left(1-n_\mathrm{s}\right)^2\left(\frac{f}{M}\right)^2<5.6\times10^{-3} $   \\ \hline
$n_{r}$ & $-0.090<n_r<0.019 $ & $-0.068<n_r<0.012 $ & $>-\frac{8}{\left(1+2N_e\right)^2}>-10^{-3} $ & $ <\frac{1}{2}\left(1-n_\mathrm{s}\right)^2<1.4\times10^{-3}  $  \\ \hline
$\Delta$ & - & - & $ \sim 5.5\times 10^{15}\left(\frac{f}{M}\right)^{1/2}GeV$ &  $ <  9 \times 10^{15}\left(\frac{f}{M}\right)^{1/2} GeV $    \\ \hline
\end{tabular}
\end{center}
\bigskip
\caption {We show expectations for the spectral indices coming from $WMAP5$ 
and $WMAP5+BAO+SN$ measurements and compare them with similar results from 
\textit{Natural Inflation} and \textit{Hybrid Natural Inflation} models. To obtain the 
numerical bounds the lowest reported value of the scalar spectral index 
$n_{\mathrm{s}}\sim 0.947$ was used. $M$ is the reduced Planck mass $M=2.44 \times 10^{18} GeV$ and $N_e$ is the number of observable e-folds.}
\label{table:1}
\end{table}

\section{A simple model of hybrid natural inflation}

We first consider a simple (non-supersymmetric) model which has a non
Abelian discrete symmetry, the semi direct product group $D_{4}$ with
generators given by the semi-direct product group elements of $Z_{2}\ltimes
Z_{2}^{\prime }.$ This will illustrate the mechanism leading to hybrid
natural inflation but will require a parameter to be chosen very small. In
the next section we will construct a supersymmetric version of the model
which avoids the need to have such a small parameter.

$D_{4}$ has $5$ irreducible representations $1^{++},1^{+-},1^{-+},1^{--},$
and $2.$ The fields in the model consist of a doublet of real scalar fields $%
\varphi =\left( 
\begin{array}{c}
\varphi _{1} \\ 
\varphi _{2}%
\end{array}%
\right) $ and two real singlet fields $\chi _{1,2}$ transforming as $1^{-+}$
and $1^{--}$ respectively$.$ Under the $Z_{2}\ltimes Z_{2}^{\prime }$ group
the fields $\varphi _{i}$ and $\chi _{1,2}$ transform as shown in Table \ref%
{Tabledoublet copy(1)}.
The leading order independent terms in the scalar potential consistent with this symmetry are given by%
\begin{equation}
V\left( \varphi ,\chi \right) =c-m_{\varphi }^{2}\varphi ^{2}+\lambda
_{1}\left( \varphi ^{2}\right) ^{2}+\sum\limits_{i=1}^{2}m_{\chi
_{i}}^{2}\chi _{i}^{2}+\lambda _{2}\sum_{i=1}^{2}\varphi _{i}^{4}-\lambda
_{3}\varphi _{1}\varphi _{2}\chi _{1}\chi _{2}+O\left( \chi ^{4}\right) ,
\end{equation}%
where $c$ is a constant and $\varphi ^{2}=\varphi _{1}^{2}+\varphi _{2}^{2}$
The first four terms are invariant under a larger $SO(2)$ $(U(1))$
symmetry, but this is broken to $D_{4}$ by the fifth and sixth terms.
For clarity, we have not included the terms $\phi^2 \chi_{1,2}^2$ as these do 
not break the $U(1)$ symmetry and do not affect the nature of  the 
inflation produced by the potential. After
symmetry breaking $\varphi ^{2}$ acquires a vacuum expectation value $(vev)$ 
$v^{2}$. Treating the fifth (and sixth) term as a perturbation it is
convenient to use the following parameterisation 
\begin{equation}
\varphi =(\rho +v)\left( 
\begin{matrix}
\cos \left( \frac{\phi }{v}\right) \cr\sin \left( \frac{\phi }{v}\right) \cr%
\end{matrix}%
\right) .
\end{equation}%
Here $\phi $ is the pseudo Goldstone boson associated with the spontaneous
symmetry breaking of the approximate $U(1)$ symmetry. It will be identified
with the inflaton. The field $\rho $ is massive and does not play a role in
the inflationary era. The potential for the $\phi $ and $\chi $ fields has
the form 
\begin{equation}
V\left( \phi ,\chi \right) =c^{\prime }+64\lambda _{2}f^{4}\cos \left( \frac{%
\phi }{f}\right) +\sum\limits_{i=1}^{2}m_{\chi _{i}}^{2}\chi
_{i}^{2}-8\lambda _{3}f^{2}\sin \left( \frac{\phi }{2f}\right) \chi _{1}\chi
_{2}+O\left( \chi ^{4}\right) ,
\end{equation}%
where $v=4f$ and $c^{\prime }\equiv c-16 m_{\varphi }^{2} f^2+256 \lambda _{1}f^4+192 \lambda _{2}f^4$ is a constant. One may see that$\footnote{%
To simplfy the presentation we take $m_{\chi _{1,2}}=m_{\chi }.$ Note that the
hybrid mechanism applies also to the case the masses are not equal.}$,
starting from $\phi \simeq 0$ with $8\lambda _{3}f^{2}\sin \left( \frac{\phi 
}{2f}\right) <m_{\chi }^{2}$, $\phi $ will \textquotedblleft
roll\textquotedblright\ until the term quadratic in the $\chi $ fields
drives $\chi _{i\text{ }}$to acquire $vevs$. If the stabilising terms,
quartic in $\chi ,$ have $O(1)$ coefficients the minimum will be at $\chi
_{i}\simeq m_{\chi }$ and so the change in potential energy following from
the $\chi $ roll is of $O\left( m_{\chi }^{4}\right) .$ In order to have
zero (or very small) cosmological constant after inflation it is necessary
to tune the parameters such that $V(\phi _{min},\chi _{min})\simeq 0$ (in
the absence of a solution to the cosmological constant problem such fine
tuning is an ingredient of all inflationary models). In this case such
tuning requires the choice $c^{\prime }=O\left( m_{\chi }^{4}\right) .$ With
this the form of the inflationary potential \textit{during} inflation near
the scale $\phi =\phi _{\mathrm{H}}$ at which the density perturbations are
produced is given by%
\begin{equation}
V\left( \phi ,0\right) =V_{0}\left( 1+a\cos \left( \frac{\phi }{f}\right)
\right) ,  \label{hni1}
\end{equation}%
where $V_{0}\equiv c^{\prime }=O\left( m_{\chi }^{4}\right) $ and $a\equiv\left( \frac{64\lambda
_{2}f^{4}}{V_0}\right) =O\left( \frac{64\lambda
_{2}f^{4}}{m_{\chi }^{4}}\right) .$ The end of inflation occurs when $\phi
=\phi _{e}$ where 
\begin{equation}
8\lambda _{3}f^{2}\sin \left( \frac{\phi _{e}}{2f}\right) \simeq m_{\chi
}^{2}.  \label{end}
\end{equation}%
\begin{table}[tbp] \centering 
\begin{tabular}{|l|l|l|}
\hline
& $\mathbf{Z}_{2}$ & $\mathbf{Z}_{2}^{{\prime }}$ \\ \hline
$\varphi _{1}$ & $\varphi _{2}$ & $\varphi _{1}$ \\ 
$\varphi _{2}$ & $\varphi _{1}$ & $-\varphi _{2}$ \\ \hline
$\chi _{1}$ & $-\chi _{1}$ & $\chi _{1}$ \\ \hline
$\chi _{2}$ & $-\chi _{2}$ & $-\chi _{2}$ \\ \hline
\end{tabular}%
\caption{The table shows the discrete transformations of the field
components under the $Z_{2}$ and $Z_{2}^{\prime}$ groups. The semi-direct
product of $Z_{2}$ with $Z_{2}^{\prime}$ results in the non Abelian dihedral
group $D_4$ and the breaking of this symmetry leads to the natural hybrid
inflation models discussed below.} \label{Tabledoublet copy(1)}%
\end{table}
The freedom in choosing the free parameters $\lambda _{2},\lambda _{3}$ and $%
m_{\chi }$ means one can adjust $V_{0\text{, }}\phi _{e}$ and $a$ at will
for any choice of $f.$ Given this it is reasonable to treat $\phi _{e}$ as a
parameter in place of some combination of $m_{\chi }$ and $\lambda _{3}$.
Further, since $\phi _{\mathrm{H}}$ is directly determined by $\phi _{e}$
and the number of $e-$folds of inflation, $N,$ between $\phi _{\mathrm{H}}$
and $\phi _{e},$ it is even more convenient to treat $\phi _{\mathrm{H}}$ as
the independent parameter. In this case the hybrid natural inflation is
completely parameterised by the potential Eq.$\left( \ref{hni1}\right) $ for 
$\phi \simeq \phi _{\mathrm{H}},$ with $N$ subsequent $e-$folds of inflation$%
.$ This is the form of the hybrid natural inflationary potential that was
analysed above. Note that the scale of inflation is given by $V_{0}=O\left(
m_{\chi }^{4}\right) $ and, since $m_{\chi }$ is a free parameter, it can be
made arbitrarily small $\footnote{%
However radiative corrections will typically drive $m_{\chi}$ to be close to the Planck scale -the usual ''hirarchy problem``. In the next section we construct a supersymmetric model that avoids this problem.}$. This represents a significant difference compared to
the case of natural inflation in which the scale is constrained to be high 
\cite{Freese:1991rb}.

It remains to check whether acceptable inflation occurs for reasonable
values of the parameters. It is easy to generate a sufficient number of $e-$%
folds between $\phi _{\mathrm{H}}$ and $\phi _{e}$, consistent with our
assumption that $\phi \ll f.$ This follows because $N\propto e^{Ht}$ while $%
\phi _{\mathrm{e}}=\phi _{\mathrm{H}}e^{\frac{m_{\phi }^{2}}{3H}t}$ and $m_{\phi }\ll
H$ due to its pseudo Goldstone nature. Thus during the last $50-60$ $e-$%
folds of inflation $\phi $ rolls only a small amount and so both $\phi _{%
\mathrm{H}}/f$ and $\phi _{e}/f$ can be small. The condition that the $\chi $
field ends inflation, $8\lambda _{3}f^{2}\sin \left( \frac{\phi }{2f}\right)
>m_{\chi }^{2},$ is readily satisfied even for the largest value of $%
m_{\chi },$ $m_{\chi }\simeq 10^{16}GeV,$ consistent with the upper bound on 
$V_{0}.$ To see this note that using Eq.(\ref{ans}) and Eq.(\ref{scale}) together with $\phi _{e} \approx \phi _{\mathrm{H}}$ Eq.(\ref{end}) is satisfied provided
\begin{equation}
\lambda _{3}\simeq \frac{A_{\mathrm{H}}}{4}\left( \frac{1-n_{\mathrm{s}}}{2}%
\right)\frac{M}{f} ,  \label{lambda3}
\end{equation}%
independent of $m_{\chi }.$
Finally, the scale of inflation is set by Eq.(\ref{scale}). Using this
together with $V_{0}\simeq m_{\chi }^{4}$, Eq.(\ref{end}) and $\phi _{%
\mathrm{H}}\simeq \phi _{e}$ one finds 
\begin{equation}
\lambda _{2}\simeq \frac{A_{\mathrm{H}}^{2}}{64}\left( \frac{1-n_{\mathrm{s}}}{2}%
\right) ^{3}\tan ^{2}\left( \frac{\phi _{\mathrm{H}}}{f}\right) .  \label{lambda2}
\end{equation}%
From this, since inflation is generated for large values of $\cos\left( \frac{%
\phi _{\mathrm{H}}}{f}\right) ,$ we see that it is necessary to choose $%
\lambda _{2}$ to be anomalously small, $\lambda _{2}\ll 10^{-14},$ certainly
a defect of this model. 
From Eq.(\ref{lambda2}) we can give an estimate for the inflaton mass $m_{\phi }^{2}\simeq 64 \lambda _{2}f^2\simeq A_{\mathrm{H}}^2\left( \frac{1-n_{\mathrm{s}}}{2}%
\right) ^{3}f^2\sim 2\times 10^{-12}f^2$.
From Eq.(\ref{lambda3}) we see that for 
$\lambda_3\leq 1,$ the scale of symmetry breaking $f$ should be $f\geq 3 \times 10^{-6}M$. 
Thus the inflationary scale Eq.(\ref{scale}) satisfies $V_0\geq \left(6 \times 10^{-6} M \tan^{1/2}\left( \frac{\phi _{\mathrm{H}}}{f}\right)\right)^4\simeq \left(10^{13} GeV \tan^{1/2}\left( \frac{\phi _{\mathrm{H}}}{f}\right)\right)^4$. However since $\phi _{\mathrm{H}}\ll f$ there is no lower bound for the inflationary scale and it can be much lower than $10^{13} GeV.$
To eliminate the need for the $\lambda_2$ fine tuning we turn
now to the construction of a supersymmetric model.

\section{A supersymmetric version of hybrid natural inflation}

It is straightforward to build a $N=1$ supersymmetric generalization of the
model. This model has the advantage that it has no anomalously small
parameters due to the additional symmetry forcing the higher order terms,
which break the approximate continuous symmetry, to be small. In addition the mass of the hybrid field is protected against radiative corrections larger than the Hubble scale during inflation and can be chosen to have any value above this scale. The field
content consists of chiral supermultiplets of the $N=1$ supersymmetry which
transform non-trivially under the $D_{4}$ non Abelian discrete symmetry. The
supermultiplets consist of a $D_{4\text{ }}$doublet $\varphi =\left( 
\begin{array}{c}
\varphi _{1} \\ 
\varphi _{2}%
\end{array}%
\right) $ and three singlet representations $\chi _{1,2}$ and $A$
transforming as $1^{-+}$, $1^{--}$ and $1^{++}$ respectively$.$ The
interactions are determined by the superpotential given by%
\begin{equation}
P=A\left( \Delta ^{2}-\lambda _{3}\frac{1}{M^{2}}\varphi _{1}\varphi
_{2}\chi _{1}\chi _{2}\right) ,
\end{equation}%
where $\Delta$ is a constant with units of mass.
These are the leading order terms if we have an additional $R-$symmetry
under which the fields $A,\varphi ,\chi $ have charges $2,-2$ and $2$
respectively. It is this $R-$symmetry that allows us to construct a model
with no anomalously small couplings. Including the effect of SUSY breaking
terms the potential has the form 
\begin{equation}
V\left( \varphi ,\chi \right) =\left\vert \Delta ^{2}-\lambda _{3}\frac{1}{%
M^{2}}\varphi _{1}\varphi _{2}\chi _{1}\chi _{2}\right\vert ^{2}+m_{\varphi
}^{2}|\varphi |^{2}+\sum\limits_{i=1}^{2}m_{\chi _{i}}^{2}\left\vert \chi
_{i}\right\vert ^{2}+\lambda _{2}\frac{m^{2}}{M_{M}^{2}}\sum_{i=1}^{2}\left%
\vert \varphi _{i}\right\vert ^{4}+O\left( \frac{m^{2}}{M^{2}}\left\vert
\chi \right\vert ^{4}\right) ,  \label{fullpot}
\end{equation}%
where we have used the same symbol for the scalar components of the chiral
supermultiplets as for the supermultiplets themselves. Here $m_{\varphi
}^{2},m_{\chi _{i}}^{2}$ and $m^{2}$ are supersymmetry breaking masses
which, during inflation, are of the order of the Hubble parameter, $H\simeq 
\frac{\Delta ^{2}}{M}.$ The last term is a $D-$term which only arise when
supersymmetry is broken and hence is proportional to the SUSY breaking
scale, $m^{2}$. It can come from radiative corrections with a messenger
field of mass $M_{M}$ in the loop.

Including radiative corrections $m_{\varphi }^{2}$ should be interpreted as
a running mass squared, $m_{\varphi }^{2}=m_{\varphi }^{2}\left( \varphi
\right) $ and these radiative corrections may readily drive it negative at
some scale $\Lambda $ below the Planck scale$.$ In this case $\varphi $ will
develop a $vev$ of order $\Lambda .$ As before we introduce a
parameterisation for $\varphi $ which is appropriate if the dominant terms
are the second and third terms of the potential that are invariant under a
larger continuous $SU(2)$ symmetry acting on the complex doublet of fields $%
\varphi .$ We treat the term proportional to $\lambda _{2}$ which breaks $%
SU(2)$ to $D_{4}$ as a perturbation. We write $\varphi $ in the form 
\begin{equation}
\varphi =e^{i\mathbf{\phi \cdot \sigma }}\left( 
\begin{array}{c}
0 \\ 
\rho +v%
\end{array}%
\right) =\frac{\rho +v}{\phi }\left( 
\begin{array}{c}
\left( \phi _{2}+i\phi _{1}\right) \sin \left( \frac{\phi }{v}\right) \\ 
\phi \cos \left( \frac{\phi }{v}\right) -i\phi _{3}\sin \left( \frac{\phi }{v%
}\right)%
\end{array}%
\right) ,
\end{equation}%
where $\sigma _{i}$ are the Pauli spin matrices, $\phi _{i}$ are the pseudo
Goldstone fields left massless by the second and third terms of the
potential and acquiring mass from the fifth term\footnote{%
For clarity, we have not included the terms $(m^2/M_M^2)|\phi|^2|\chi_{1,2}|^2$ 
as these do not break the $SU(2)$ symmetry and do not affect the nature of 
the inflation produced by the potential.} and finally $\phi ^{2}=\phi
_{1}^{2}+\phi _{2}^{2}+\phi _{3}^{2}$. The field $\rho $ acquires a mass of $%
O(m_{\varphi })$ and plays no role in the inflationary era. The potential
for the pseudo Goldstone fields $\phi _{i}$ alone is given by 
\begin{equation}
V\left( \phi \right) =\frac{m^{2}}{M^{2}_M}v^{4}\lambda _{2}\left( \left[
\left( \frac{\phi _{1}}{\phi }\right) ^{2}+\left( \frac{\phi _{2}}{\phi }%
\right) ^{2}\right] ^{2}\sin ^{4}\left( \frac{\phi }{v}\right) +\left[ \cos
^{2}\left( \frac{\phi }{v}\right) +\left( \frac{\phi _{3}}{\phi }\right)
^{2}\sin ^{2}\left( \frac{\phi }{v}\right) \right] ^{2}\right) .
\end{equation}%
We shall demonstrate that inflation occurs for small $\frac{\phi _{i}}{v}.$
In this region, for positive $\lambda _{2},$ the field $\phi _{3}$ has a
positive mass squared while $\phi _{1,2}$ have negative mass squared. Thus $%
\phi _{3}$ does not develop a $vev.$ The full potential for the fields
acquiring $vev$s then has the form%
\begin{eqnarray}
V\left( \phi ,\chi \right) &=&\Delta ^{^{\prime }4}+64\lambda _{2}\frac{m^{2}%
}{M^{2}_M}f^{4}\cos \left( \frac{\phi }{f}\right)
+\sum\limits_{i=1}^{2}m_{\chi _{i}}^{2}\left\vert \chi _{i}\right\vert ^{2}-
\notag \\
&&-8\lambda _{3}e^{i\alpha }\Delta ^{2}\frac{f^{2}}{M^{2}}\sin \left( \frac{%
\phi }{2f}\right) \chi _{1}\chi _{2}+h.c.+O\left( \frac{m^{2}}{M^{2}}\chi
^{4}\right) ,  \label{susy2}
\end{eqnarray}%
where $\phi ^{2}=\phi _{1}^{2}+\phi _{2}^{2},$ $\Delta ^{^{\prime }4}=\Delta
^{4}+16m_{\varphi }^{2}f^{2}+192\lambda _{2}%
\frac{m^{2}}{M^{2}_M}f^{4}$ and $\alpha =\tan ^{-1}\left( \frac{\phi _{1}}{%
\phi _{2}}\right) .$ Note that since the soft SUSY\ breaking masses are of $%
O(\Delta ^{\prime 2}/M)$ the last 3 terms are small for $f<M$ and $\Delta
^{\prime }\simeq \Delta .$

The condition that $\chi _{i}$ should end inflation is that in Eq.(\ref%
{susy2}) the fourth term dominates over the third, i.e.%
\begin{equation}
8\lambda _{3}\Delta ^{2}\frac{f^{2}}{M^{2}}\sin \left( \frac{\phi }{2f}%
\right) >m_{\chi }^{2}.  \label{endsusy}
\end{equation}%
Provided $\frac{\phi }{f}$ is initially small the $\chi _{i}$ fields will
initially have zero $vevs$ but once Eq.(\ref{endsusy})\ is satisfied they
will roll to their final minimum to end inflation. As may be seen from Eq.(%
\ref{fullpot}) the final minimum sets the first term to zero leaving an $%
O(m^{2}f^{2})$ contribution (small relative to $\Delta ^{4})$ that must be
cancelled by an additional constant in order to set the cosmological
constant to zero. For $f<M$ this term is small and the potential during
inflation is dominated by the $\Delta ^{4}$ term.

Thus the first two terms are of the form of the potential given in Eq.(\ref%
{hni}) with $V_{0}\simeq \Delta ^{4}$ and 
\begin{equation}
a=64\lambda _{2}\frac{m^{2}}{M_{M}^{2}}\frac{f^{4}}{\Delta ^{\prime 4}}%
\simeq 64\lambda _{2}\left( \frac{f}{M}\right) ^{4}\left( \frac{M}{M_{M}}%
\right) ^{2}.  \label{asusy}
\end{equation}%
We can now determine the parameter range that leads to acceptable inflation.
Note that, compared to the non-supersymmetric case, the scale of the
potential during inflation is still proportional to $a^{-1}$ but the
coefficient is suppressed by the factor $m^{2}/M_{M}^{2}.$ As a result the
bound of Eq.(\ref{lambda2}) is weakened by the same factor. As we will
discuss this makes it easy to get satisfactory inflation without fine tuning.

As before during the last $40-50$ $e-$folds of inflation $\phi $ rolls only
a small amount and so both $\phi _{\mathrm{H}}/f$ and $\phi _{e}/f$ can be
small. The spectral index is determined by the value of $a$ which, c.f. Eq(%
\ref{asusy}), can be adjusted for arbitrary $f$ by choosing the appropriate
mediator mass scale $M_{M}.$ From Eqs.(\ref{scale}), (\ref{asusy}) and the
relation between $a$ and $n_{s}$ we find 
\begin{eqnarray}
\lambda _{2} &=&\frac{1}{64}\left( \frac{1-n_{\mathrm{s}}}{2}\right) \left( \frac{%
M_{M}}{f}\right) ^{2} ,  \notag \\
\lambda _{3} &=&\frac{A_{\mathrm{H}}}{8}\left( \frac{1-n_{\mathrm{s}}}{2}%
\right) \left( \frac{m_{\chi }}{m}\right) ^{2}\left( \frac{M}{f}\right)  .\label{couplingbound}
\end{eqnarray}%
From this one may see that $\lambda _{2,3}$ need not be chosen unnaturally
small provided the mediator scale is significantly below the Planck scale.
Given the smallness of the first two factors in Eq.(\ref{couplingbound}) a
small $f$ is favoured, $f<10^{-4},$ so that higher order supergravity
corrections are under control. The mass scale $M_{M}$ needs to be chosen one
or two orders of magnitude larger, quite reasonable given the interpretation
of $M_{M}$ as a mediator mass. 
From Eq.(\ref{couplingbound}) we can give an estimate for the inflaton mass $m_{\phi }^{2}\simeq 64 \lambda _{2}f^2\left(\frac{m}{M_{M}}\right)^2\simeq \left( \frac{1-n_{\mathrm{s}}}{2}%
\right) m^2\sim 2\times 10^{-2}m^2$.
The scale of inflation is set by $\Delta $
and there is no constraint on it coming from Eq.(\ref{couplingbound}). As a
result this model of hybrid natural inflation provides another example of
inflationary models capable of generating very low scales of inflation \cite%
{German:2001tz}. However in this case there is no constraint on the spectral index compared to the models of \cite {German:2001tz} which required $n_{\mathrm{s}}<0.95.$ An effect of lowering the inflationary scale is a decrease of the reheat temperature due to the decays of the inflaton field $\phi$. The observed baryon asymmetry of the universe may in principle be generated after reheating through anomalous electroweak $(B-L)$-violating processes and/or the Affleck-Dine mechanism, even for reheating temperatures as low as $\sim 1$ GeV \cite%
{Davidson:2000}. In this paper we have concentrated on the inflationary era. However viable models must also have an acceptable reheat phase capable of producing the observed baryon excess.  Although a detailed discussion of this is beyond the scope of this paper it is important to point out that it is the decay of the hybrid field and not the inflaton that is responsible for reheating. This avoids the difficulties associated with the small coupling of a pseudo Goldstone boson and the reheating due to the hybrid field can be very efficient\cite{Linde:2007fr}. Furthermore, even if the scale of inflation is very low there are many suggested mechanisms capable of acceptable baryogenesis \cite{Shaposhnikov:2009zzb}.

\section{Summary and conclusions}

We have demonstrated a generalisation of \textit{%
Natural Inflation} which preserves the attractive pseudo Goldstone
origin of the inflaton capable of solving the $\eta $ problem while
implementing a hybrid end to inflation through the rapid roll of a second
scalar field. The coupling of the hybrid field to the inflaton is possible
in models which have a spontaneously broken non Abelian discrete symmetry
and the resulting form of the inflationary potential at the time the density
perturbations are produced has a very simple general form. We have analysed
the inflationary parameters resulting from hybrid natural inflation and
shown that an acceptable value for the spectral index is easy to achieve for
a range of inflationary scales from the upper Lyth bound down to arbitrarily
low scales. Over most of the allowed range of parameters the tensor to
scalar ratio and the running of the spectral index are unobservably small.
Moreover there is no need for the symmetry breaking scale to be super
Planckian so it is easy to avoid large gravitational corrections.

Using a simple $D_{4}$ non Abelian discrete symmetry we constructed two very
simple hybrid natural inflationary models, one non supersymmetric and one
supersymmetric. In the non supersymmetric case it was necessary to choose
the coefficient of the leading term breaking the approximate continuous
symmetry to be anomalously small. The second example implements the non
Abelian discrete symmetry in a supersymmetric theory. The additional
constraints of supersymmetry allows for the construction of a viable model
with no parameter chosen to be anomalously small and the scale of inflation
to be arbitrary.

The $D_{4}$ based examples provide simple realisations of hybrid natural
inflation but there are many other possible realisations following from
alternative non Abelian discrete symmetries. That such symmetries may occur
in nature is perhaps made more plausible by the fact that they readily occur
in compactified string theories. Irrespective of the details of the discrete
group we expect the form of the inflationary potential at the time the
density perturbations are produced to have the universal form presented
above and so the expectations for the inflationary parameters will have the
structure discussed.


\begin{thebibliography}{99}
\bibitem{Freese:1990rb} K.~Freese, J.~A.~Frieman and A.~V.~Olinto, 
Phys.\ Rev.\ Lett.\ \textbf{65} (1990) 3233; 
F.~C.~Adams, J.~R.~Bond, K.~Freese, J.~A.~Frieman and A.~V.~Olinto, 
Phys.\ Rev.\ D \textbf{47} (1993) 426 [arXiv:hep-ph/9207245]. 

\bibitem{Freese:1991rb} K.~Freese, C.~Savage and W.~H.~Kinney, 
Int.\ J.\ Mod.\ Phys.\ D \textbf{16} (2008) 2573 [arXiv:0802.0227 [hep-ph]]. 


\bibitem{Shaun:2008} S.~Hotchkiss, G.~Germ\'an, G.~G.~Ross and S.~Sarkar, 
JCAP \textbf{0810} (2008) 15 [arXiv:0804.2634 [astro-ph]]. 

\bibitem{Kobayashi:2006wq} T.~Kobayashi, H.~P.~Nilles, F.~Ploger, S.~Raby
and M.~Ratz, 
Nucl.\ Phys.\ B \textbf{768} (2007) 135 [arXiv:hep-ph/0611020]. 

\bibitem{Linde:1994} A.~Linde, Phys.\ Rev.\ D \textbf{49} (1994)748.  Some general references for hybrid inflation are: D.~H.~Lyth and A.~Riotto,
  Phys.\ Rept.\  {\bf 314} (1999) 1
  [arXiv:hep-ph/9807278];
  A.~D.~Linde,
  Lect.\ Notes Phys.\  {\bf 738} (2008) 1
  [arXiv:0705.0164 [hep-th]];
  M.~Shaposhnikov,
  J.\ Phys.\ Conf.\ Ser.\  {\bf 171} (2009) 012005.

\bibitem{Stewart:2000pa} E.~D.~Stewart and J.~D.~Cohn, 
Phys.\ Rev.\ D \textbf{63} (2001) 083519 [arXiv:hep-ph/0002214]. 


\bibitem{cohn:2000} J.~D.~Cohn and E.~D.~Stewart, 
Phys.\ Lett.\ B \textbf{475} (2000) 231 [arXiv:hep-ph/0001333]. 

\bibitem{Antusch:2008gw} S.~Antusch, S.~F.~King, M.~Malinsky,
L.~Velasco-Sevilla and I.~Zavala, 
Phys.\ Lett.\ B \textbf{666} (2008) 176 [arXiv:0805.0325 [hep-ph]]. 


\bibitem{Liddle:2000cg} A.~R.~Liddle and D.~H.~Lyth, \textit{Cosmological
Inflation and Large-Scale Structure}, 
Cambridge University Press, (2000).

\bibitem{Komatsu:2008} E. Komatsu, et al., ApJS, \textbf{180} (2009)330, arXiv:0803.0547[astro-ph].

\bibitem{Dunkley:2008} J. Dunkley, et al., ApJS, \textbf{180} (2009) 306, arXiv:0803.0586[astro-ph].

\bibitem{Percival:2007} W.~J.~Percival et al., Mon.~Not.~Roy.~Astron.~Soc.~\textbf{381} (2007) 1053, arXiv:0705.3323[astro-ph].

\bibitem{Riess:2004} A.~G.~Riess et al., Astrophys.~J.~\textbf{607} (2004) 665, arXiv:0402512[astro-ph]; A.~G.~Riess et al., Astrophys.~J.~\textbf{659} (2007) 98, arXiv:0611572[astro-ph]. 

\bibitem{Astier:2006} P.~Astier et al., Astron.~Astrophys.~\textbf{447} (2006) 31, arXiv:0510447[astro-ph]. 

\bibitem{Woods-Vasey:2006} W.~M.~Woods-Vasey et al., Astrophys.~J.~\textbf{666} (2007) 694, arXiv:0701041[astro-ph].


\bibitem{Lyth:1996im} D.~H.~Lyth, 
Phys.\ Rev.\ Lett.\ \textbf{78} (1997) 1861 [arXiv:hep-ph/9606387]. 


\bibitem{German:2001tz} G.~Germ\'an, G.~G.~Ross and S.~Sarkar, 
Nucl.\ Phys.\ B \textbf{608} (2001) 423 [arXiv:hep-ph/0103243]; G.~Germ\'an, G.~G.~Ross and S.~Sarkar, Phys.\ Lett.\ \textbf{B469} (1999) 46-54,1999 [arXiv:hep-ph/9908380].


\bibitem{Davidson:2000} See, e.g., S.~Davidson, M.~Losada, A.~Riotto,  
Phys.\ Rev.\ Lett.\ \textbf{84} (2000) 4284. 
\bibitem{Linde:2007fr}
  A.~D.~Linde,
  Lect.\ Notes Phys.\  {\bf 738} (2008) 1
  [arXiv:0705.0164 [hep-th]].

\bibitem{Shaposhnikov:2009zzb}
  M.~Shaposhnikov,
  J.\ Phys.\ Conf.\ Ser.\  {\bf 171} (2009) 012005.


\end{thebibliography}
\end{document}